\documentclass{aastex}
\usepackage{spr-astr-addons}
\usepackage{url}

\begin{document}

\title{Effects of Scalar-Tensor-Vector Gravity on relativistic jets}
\shorttitle{Effects of STVG on relativistic jets}
\shortauthors{F.G. Lopez Armengol, G.E. Romero}
\author{Federico G. Lopez Armengol} \and \author{Gustavo E. Romero\altaffilmark{1}}
\affil{Instituto Argentino de Radioastronom\'ia (CCT-La Plata, CONICET; CICPBA), C.C. No. 5, 1894,Villa Elisa, Argentina}
\email{flopezar@iar.unlp.edu.ar}

\altaffiltext{1}{Also at Facultad de Ciencias Astron\'omicas y Geof\'isicas, Universidad Nacional de La Plata, Paseo del Bosque s/n, 1900 La Plata, Buenos Aires, Argentina.}

\begin{abstract}
   Scalar-Tensor-Vector Gravity (STVG) is a theory that does not require dark components to describe astrophysical data. We aim at constraining the free parameters of STVG based on recent observations of the jet in M87. We derive the equations of motion for particles in STVG-Kerr spacetime, we develop a numerical code that integrates such equations, and apply it to the jet of M87. We find that STVG deviates from GR and we set new upper limits for the free parameters of the former. We conclude that STVG is not contradicted by the observational data of M87, and may help to explain jet formation.
\end{abstract}


\section{Introduction}
\label{s:intro}

	It has been nearly eighty years since the publication of \citet{babcock1939} noticing the discrepancies between observed galactic rotation curves and theoretical predictions. Mainstream solutions to this problem have involved the postulation of dark matter. However, every experiment aimed at measuring properties of this kind of matter has failed \citep{aprile2012,lux2014,agnese2014}.  In this context, alternative solutions involving modifications of fundamental physical laws deserve some attention.\par

	\citet{milgrom1983} was the first to account for astrophysical phenomena without dark matter, introducing the MOdified Newtonian Dynamics (MOND) theory. In the subsequent years, several relativistic theories whose weak field limit coincides with MOND were formulated. See \citet{famaey2012} for a review of MOND predictions and its relativistic extensions. Motivated by problems on such extensions, \citet{moffat2006} postulated the Scalar-Tensor-Vector Gravity theory (STVG), also referred as MOdified Gravity (MOG) in the literature.\par

	In STVG, the gravitational coupling constant $G$ is reified to a scalar field whose numerical value exceeds Newton's constant $G_{\textrm{N}}$. This assumption serves to correctly describe galaxy rotation curves \citep{brownstein2006}, cluster dynamics \linebreak \citep{moffat2014}, and Bullet Cluster phenomena \citep{brownstein2007}, without requiring the existence of dark matter. In order to counteract the enhanced gravitational coupling constant close to the gravitational source, Moffat included a repulsive vector field of short range. In this way, Newton's gravitational constant can be retrieved and STVG coincides with General Relativity (GR), for instance, in the Solar System. The vector field can also mimic the effects of dark matter in the growth of cosmological structures \citep{shojai2017}. However, according to \citet{jamali2016}, any extra field can play the role of dark energy, so the theory still requires a non-vanishing cosmological constant. Further studies of STVG include the gravitational Jeans instability \citep{roshan2014}, the structure of neutron stars \citep{lopezarmengol2017}, the emission of accretion disks around black holes \citep{perez2017}, and the stability of galactic disks, where the theory proved to play a similar stabilizing effect as dark matter halos \citep{ghafourian2017}.\par
	
	The interplay between enhanced attraction and repulsion can be seen from the radial acceleration of a test particle in the weak field, static, spherically symmetric, and constant scalar field approximation \citep{moffat2006}:
\begin{equation}
\label{eq:weakfield}
	a(r)=-\frac{G_\textrm{N}(1+\alpha) M}{r^2} + \frac{G_{\textrm{N}}\alpha M}{r^2} e^{-m_{\phi}r}  (1+m_{\phi}r),
\end{equation}
where $M$ denotes the gravitational mass source, $r$ the distance from it, $G_{\textrm{N}}$ is Newton's gravitational constant, and $\alpha,m_{\phi}$ are free parameters of the theory. The first term in Eq. (\ref{eq:weakfield}) results in an enhanced attraction, quantified by $G_{\infty}=G_{\textrm{N}}(1+\alpha)$, and prevails at $r\rightarrow \infty$ . This term describes correctly galaxy rotation curves, light bending phenomena, and cosmological data without dark matter. The second term represents gravitational repulsion and is important when $m_{\phi}r<<1$. This short range force cancels the increase of $G_{\infty}$ given by $\alpha$ and retrieves $G_{\textrm{N}}$ as the gravitational coupling constant in the vicinity of the gravitational source.\par

	According to Eq. (\ref{eq:weakfield}), the differences between STVG and GR manifest far from the gravitational source, where phenomena related to dark matter use to happen. However, we should notice that such equation is based on several assumptions that may fail in the strong field regime.\par
	
	The purpose of this work is to compare GR and STVG close to the gravitational source, investigate whether they differ on shorter scales as well, and constrain the free parameters of STVG using new high resolution radio observations of the relativistic jet of the nearby galaxy M87. Our objects of study are rotating black holes and the trajectories of test particles close to them.\par

	Actually, we expect peculiar deviations. The repulsive force that counteracts the enhanced attraction is led by a vector field, and vector forces are not restricted to the radial direction; they have azimuthal or polar components instead, like the Lorentz force in Electromagnetism (EM). We expect STVG to predict novel gravitational \textit{Lorentz-like} effects, completely absent in GR.\par

	Relativistic jets, launched from the surroundings of supermassive black holes in active galactic nuclei (AGNs), should be sensitive to such Lorentz-like effects. This is because the launching region is near the event horizon, where the strong field is important, and because of the highly relativistic velocities involved. In the case of nearby sources such effects might be observationally detectable. Particularly, the extragalactic jet of the giant elliptical galaxy M87 (a.k.a Virgo A, NGV4486, and 3C274) has been resolved up to 100 gravitational radii using Very Long Baseline Interferometry (VLBI) \citep{mertens2016}. This, along with constraints on the size of M87 supermassive black hole (hereafter M87*) from mm-VLBI observations \citep{broderick2015}, might provide a unique scenario to test some predictions of STVG.\par

	Our work is organized as follows: in Section \ref{s:stvgaction} we present the action and field equations of STVG, along with certain simplifications. In Section \ref{s:stvgkerr} we describe STVG-Kerr black hole and spacetime, and derive the equations of motion for test particles. Then, in Section \ref{s:numerical} we explain a numerical method developed to integrate such equations for particles in a relativistic jet. Section \ref{s:results} is devoted to our main results for  the case of M87*, and consequent constraints on the free parameters of STVG. In Section \ref{s:discussion} we discuss the applicability of the theory to the formation of relativistic jets, and in Section \ref{s:conclusions} we present our main conclusions.

\section{STVG action and field equations}
\label{s:stvgaction}

	STVG action reads\footnote{Compared with the original action in \citet{moffat2006}, we drop the cosmological constant term because its effects are locally negligible. We also ignore the scalar field $\omega$ and set the potential $W(\phi)=0$ as suggested by \citet{moffat2013,moffat2006}, respectively.}:
\begin{equation}
\label{eq:totalaction}	
	S=S_{\textrm{GR}}+S_{\phi}+S_{\textrm{S}}+S_{\textrm{M}},
\end{equation}
where
\begin{equation}
S_{\textrm{GR}}= \frac{1}{16 \pi } \int d^4x \sqrt{-g} \frac{1}{G} R,
\end{equation}
\begin{equation}
S_{\phi}=   -\int d^4x \sqrt{-g} \left( \frac{1}{4} B^{\mu \nu} B_{\mu \nu} - \frac{1}{2} m_{\phi}^2 \phi^\mu \phi_\mu \right),
\end{equation}
\begin{eqnarray}
S_{\textrm{S}}=&&  \int d^4x \sqrt{-g} \left[ \frac{1}{G^3} \left(\frac{1}{2} g^{\mu \nu} \nabla_\mu G \nabla_\nu G - V(G) \right)+\right. \nonumber \\ 
+&&\left. \frac{1}{G m_{\phi}^2} \left(\frac{1}{2} g^{\mu \nu} \nabla_\mu m_{\phi} \nabla_\nu m_{\phi} - V(m_{\phi}) \right)\right].
\end{eqnarray}
Here, $g_{\mu \nu}$ denotes the spacetime metric, $R$ the Ricci scalar, and $\nabla_{\mu}$ the covariant derivative; $\phi^{\mu}$ denotes a Proca-type massive vector field, $m_{\phi}$ its mass, and $B_{\mu \nu}=\partial_{\mu} \phi_{\nu} - \partial_{\nu} \phi_{\mu}$; $V(G)$ and $V(m)$ denote possible potentials for the scalar fields $G(x)$ and $m_{\phi}(x)$, respectively. We adopt the metric signature $\eta_{\mu \nu}=\textrm{diag(}-1,1,1,1\textrm{)}$ and natural units. The term $S_{\textrm{M}}$ in the action refers to possible matter sources.\par

	We take certain simplifications into account: we neglect the mass $m_{\phi}$ of the vector field because its effects manifest at kiloparsecs from the source, and our region of interest is contained within sub-parsec scales. Physically, this means that we are not considering the decay of the Yukawa-type force. The same approximation has been made in \citet{moffat2015,hussain2015}.\par

	Further, we approximate the scalar field $G$ as a constant and adopt the same prescription as \citet{moffat2006}:
\begin{equation}
\label{eq:G}
	G_{\infty}=G_{\textrm{N}}(1+\alpha),
\end{equation}
where $\alpha$ is a free parameter whose value we sample.\par

	Lastly, we nullify the matter action term $S_{\textrm{M}}$ because we study the vacuum spacetime of a rotating black hole.\par

	The simplified action takes the form:
\begin{equation}
\label{eq:simpleaction}
	S=\int d^4x \sqrt{-g}\left[ \frac{1}{16 \pi G_{\infty}}  R - \frac{1}{4} B^{\mu \nu} B_{\mu \nu} \right],
\end{equation}
which formally resembles the Einstein-Maxwell action, and suggests the existence of gravitational Lorentz-like effects in STVG. \par

	By varying the simplified action (\ref{eq:simpleaction}) with respect to the metric $g^{\mu \nu}$ we obtain:
\begin{equation}
\label{eq:metric}
	G_{\mu \nu} =  8\pi G_{\infty} T^{\phi}_{\mu \nu} ,
\end{equation}
where $G_{\mu \nu}$ denotes the Einstein tensor and 
\begin{equation}
	\label{eq:energymomentum2}
	T^{{\phi}}_{\mu \nu} = -\frac{2}{\sqrt{-g}} \frac{\delta S_{\phi}}{\delta g^{\mu \nu}}= \left({B_{\mu}}^{\alpha} B_{\nu \alpha} - g_{\mu \nu} \frac{1}{4} B^{\rho \sigma} B_{\rho \sigma}		\right).
	\end{equation}\par

	Furthermore, varying the action (\ref{eq:simpleaction}) with respect to the vector field $\phi_{\mu}$ yields:
\begin{equation}
\label{eq:vector}
	\nabla_{\nu} B^{\nu \mu} = 0.
\end{equation}\par

	Finally, the equations of motion for a test particle in coordinates $x^{\mu}$ are given by:
\begin{equation}
	\label{eq:motion}
	\left(\frac{d^2 x^{\mu}}{d \tau^2} + \Gamma^{\mu}_{\alpha \beta} \frac{dx^{\alpha}}{d\tau} \frac{dx^{\beta}}{d\tau} \right)=\frac{q}{m} {B^{\mu}}_{\nu} \frac{dx^{\nu}}{d\tau},
\end{equation}
where $\tau$ denotes the particle proper time, and $q$ the coupling constant with the vector field. We define the parameter:
\begin{equation}
\label{eq:kappa}
\kappa=\frac{q}{m},
\end{equation}
whose value we will sample, along with $\alpha$.

\section{STVG-Kerr spacetime}
\label{s:stvgkerr}

	\citet{moffat2015} investigated the STVG-Kerr spacetime. This is the vacuum and axially symmetric solution to the metric field equations (\ref{eq:metric}), for a body with mass $M$ and spin per unit mass $a$. In Boyer-Lindquist coordinates, it reads:
\begin{eqnarray}
\label{eq:kerrmetric}
	ds^2=&&-\frac{\Delta}{\rho^2} \left[d(ct) -\frac{a \sin^2 \theta}{c} d \phi\right]^2 + \nonumber \\*
	&&+\frac{\sin^2 \theta}{\rho^2} \left[\left(r^2+\frac{a^2}{c^2}\right) d\phi - \frac{a}{c} d(ct) \right]^2 + \nonumber \\*
	&&+ \frac{\rho^2}{\Delta} dr^2 + \rho^2 d\theta^2,
\end{eqnarray}
where 
\begin{equation}
	\Delta=r^2-\frac{2G_{\infty}M}{c^2}r+\frac{a^2}{c^2}+ \frac{G_{\textrm{N}} Q^2}{c^4},
\end{equation}
\begin{equation}
	\rho^2=r^2+\frac{a^2}{c^2} \cos^2 \theta,
\end{equation}
\begin{equation}
	Q=\sqrt{\alpha G_{\textrm{N}}} M.
\end{equation}\par

	The black hole geometry (\ref{eq:kerrmetric}) presents two horizons, given by the roots of $\Delta=0$:
\begin{equation}
\label{eq:eventhorizon}
	r_{\pm}=\frac{G_{\infty}M}{c^2}\left(1\pm\sqrt{1-\frac{c^2 a^2}{G_{\infty}^2M^2}-\frac{\alpha}{1+\alpha}}\right),
\end{equation}
and an ergosphere determined by the roots of $g_{00}=0$:
\begin{equation}
\label{eq:ergosphere}
	r_{\textrm{E}}=\frac{G_{\infty}M}{c^2}\left(1\pm\sqrt{1-\frac{c^2 a^2\cos^2\theta}{G_{\infty}^2M^2}-\frac{\alpha}{1+\alpha}}\right).
\end{equation}
Furthermore, the spacetime possesses a ring singularity, given by the roots of $\rho=0$.\par

	Such geometrical features change with $\alpha$. In particular, the external event horizon of a rotating black hole in STVG is bigger than in GR. In Section \ref{s:results} we make use of this fact to constrain the value of $\alpha$ from mm-VLBI observations of M87*.\par

	Having described the STVG-Kerr spacetime, we turn our attention to the vector field $\boldsymbol{\phi}$. The vector field equation (\ref{eq:vector}) for the black hole geometry (\ref{eq:kerrmetric}) has been studied exhaustively in the context of Einstein-Maxwell theory (see, for instance, \citet{misner1973}). Adapting such results to STVG, we find:
\begin{eqnarray}
\label{eq:bmunu}
	\boldsymbol{B}=\frac{Q}{c\rho^4} \left(r^2-a^2 \cos^2\theta\right) \boldsymbol{d}r \wedge \left[ \boldsymbol{d}t -\frac{a}{c} \sin^2\theta \boldsymbol{d} \phi \right] + \nonumber \\*
	+\frac{2Qa}{c^2\rho^4} r \cos\theta \sin\theta \boldsymbol{d}\theta \wedge \left[ \left(r^2+\frac{a^2}{c^2}\right)\boldsymbol{d}\phi-\frac{a}{c} \boldsymbol{d}t\right],
\end{eqnarray}
that corresponds to the vector potential:
\begin{equation}
\label{eq:phimu}
	\boldsymbol{ \phi }=-\frac{Qr}{\rho^2}\left(\mathbf{d}t-a\sin^2\theta \mathbf{d}\phi\right).
\end{equation}\par

\begin{figure}
	\centering
  \includegraphics[angle=270,width=.9\columnwidth]{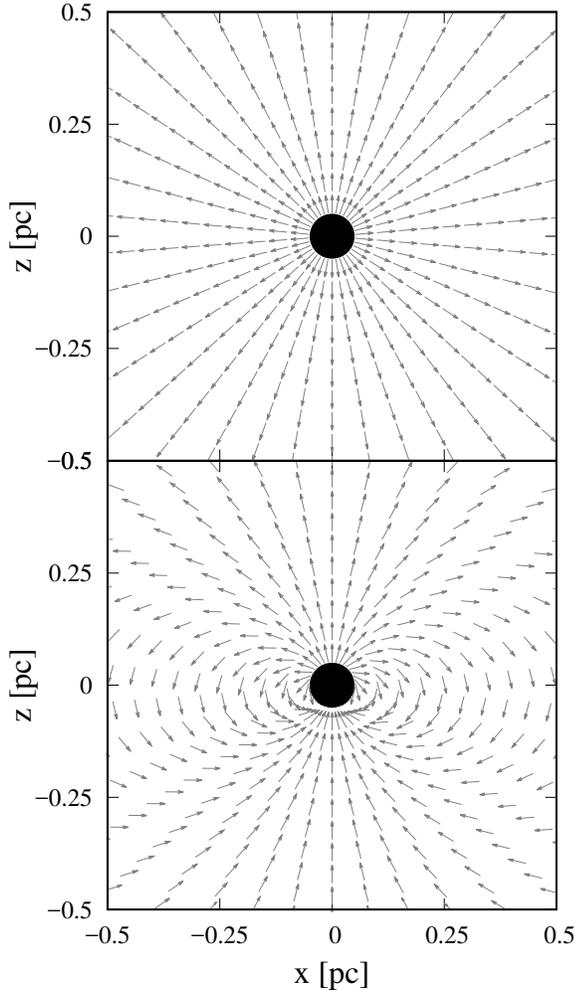}
  \caption{Vector maps of $B^{\mu\nu}$ on  Kerr-Schild $x-z$ plane. The field is generated by a supermassive black hole with mass $M=6\times10^{9}M_{\odot}$, and angular momentum $a=0.9 G_{\textrm{N}}M/c$. The field lines are normalized. \textit{Top:} Gravito-electric components $B^{0i}$. These radial components generate a repulsive force that counteracts the enforced attraction, and retrieves Newton gravitational law on the right scale. \textit{Bottom:} Gravito-mangetic components $B^{ij}$. The field lines have the familiar disposition of a magnetic dipole generated by a rotating charge. The effects of these components involve novel predictions of STVG}
	\label{fig1}
\end{figure}

	Following Moffat's idea, the \emph{gravito-electrical} components $B^{0i}$ counteract the enhanced attraction. However, \emph{gravito-magnetic} components $B^{ij}$ give raise to azimuthal and polar forces, completely absent in GR. In Fig. \ref{fig1} we map the latter components.\par

	We then proceed to study the trajectory of a test particle with mass $m$ in STVG-Kerr spacetime. The equations of motion (\ref{eq:motion}), for the geometry (\ref{eq:kerrmetric}) and tensor field (\ref{eq:bmunu}), has been treated in the context of Einstein-Maxwell theory (see \citet{carter1968,misner1973}). Making use of these results, we obtain a system of first order differential equations:
\begin{equation}
	\label{eq:drdlambda}
	\rho^2 \frac{dr}{d\lambda}=\pm \sqrt{R(r)},
\end{equation}
\begin{equation}
	\label{eq:dthetadlambda}
	\rho^2 \frac{d\theta}{d\lambda} = \pm \sqrt{\Theta (\theta)},
\end{equation}
\begin{equation}
	\label{eq:dphidlambda}
	\rho^2 \frac{d\phi}{d\lambda} =-\left(\frac{a E}{c^2} - \frac{a}{\sin^2\theta}\right) + \frac{a P(r)}{\Delta(r) c^2},
\end{equation}
\begin{equation}
	\label{eq:dtdlambda}
	\rho^2 \frac{dt}{d\lambda}=-\frac{a\sin^2\theta}{c^2}\left(\frac{a E}{c^2} - \frac{L}{\sin^2\theta}\right) + \left(r^2+\frac{a^2}{c^2}\right) \frac{P(r)}{\Delta(r) c^2},
\end{equation}
where $\lambda=\tau/m$, $E$ stands for the energy of the test particle, and $L$ for its angular momentum around the symmetry axis. Both $E$ and $L$ are constants of motion. Further, we have the functions:
\begin{equation}
	R(r)=\frac{P^2(r)}{c^2} - \Delta(r)\left(m^2r^2c^2+\mathcal{K}\right),
\end{equation}
\begin{equation}
	\Theta(\theta)=\mathcal{Q}-\cos^2\theta\left[a^2\left(m^2-\frac{E^2}{c^4}\right) + \frac{L^2}{\sin^2\theta}\right],
\end{equation}
\begin{equation}
	P(r)=E\left(r^2+\frac{a^2}{c^2}\right) - aL- q Q r,
\end{equation}
where $\mathcal{K}$ is Carter's constant of motion, and $\mathcal{Q}$ a particular combination of constants. We present expressions for the latter in the following section.\par

\section{Numerical treatment}
\label{s:numerical}

	We develop a numerical code that integrates the system of differential equations (\ref{eq:drdlambda})-(\ref{eq:dtdlambda}) for a particle in a relativistic jet. The input variables of the code are observational parameters of the astrophysical jet:

\begin{enumerate}
	\item We fix the spacetime geometry and fields by setting $M,\ a$, and $\alpha$. The values of $\alpha$ are obtained through the parameter $M_0$, studied extensively by Moffat and collaborators \citep{brownstein2006,brownstein2007}. Both parameters are related by:
	\begin{equation}
		\alpha=\sqrt{\frac{M_0}{M}}.
	\end{equation}

	\item We set the intrinsic properties of the particle $m, \ \kappa$, and its initial position $(r_0, \ \theta_0, \ \phi_0)$. Without loss of generality, we set $\phi_0=0$.

	\item Now we focus on the initial values for $p_{\mu}=dx_{\mu}/d\lambda$. Since we are interested in azimuthal effects given by gravito-magnetic forces, we set initially:
	\begin{equation}
		p_{\phi}=0. 
	\end{equation}
	
	\item For the $t$-component, we have:
	\begin{equation}
		\label{eq:pt}
		p_t=-mc\gamma\sqrt{g_{tt}},
	\end{equation}
where $\gamma$ is the local Lorentz factor of the particle. 

	\item The intial components $p_r, \ p_{\theta}$ require further steps because they depend on the ejection angle $\theta_{\mathrm{ej}}$ between the initial velocity and the $z$-axis. Since the $z$-axis is well defined in Kerr-Schild coordinates, first we have to solve the system of non-linear equations for the initial Kerr-Schild momentum components $\tilde{p}_x, \ \tilde{p}_z$:
	\begin{eqnarray}
		\cos\theta_{\mathrm{ej}}=\frac{\tilde{p}^i z^j g_{ij}}{\sqrt{\tilde{p}^i\tilde{p}_i} \sqrt{z^i z_i}}= \nonumber \\*
		= \frac{\tilde{p}^x z^x g_{xx}+\tilde{p}^x z^z g_{xz}+\tilde{p}^z z^z g_{zz}}{\sqrt{\left(\tilde{p}^x\right)^2 g_{xx}+2\tilde{p}^x \tilde{p}^z g_{xz}+\left(\tilde{p}^z\right)^2 g_{zz}}\sqrt{g_{zz}}},
	\end{eqnarray}
	\begin{eqnarray}
		\tilde{p}^{\mu}\tilde{p}_{\mu}= \left(p_t\right)^2 g^{tt}+2\tilde{p}^x p_t +2\tilde{p}^z p_t +\left(\tilde{p}^x\right)^2 g_{xx} +\nonumber \\*
			+2\tilde{p}^x \tilde{p}^z g_{xz}+\left(\tilde{p}^z\right)^2 g_{zz}=-m^2 c^2,
	\end{eqnarray}
where $z^i=(0,0,1)$. We set $\tilde{p}_y=0$ in consistence with $p_{\phi}=0$, and we take $p_t$ from Eq. (\ref{eq:pt}). We solve the non-linear system of equations applying a Newton-Raphson subroutine from \citet{press1992}. After finding the initial components $\tilde{p}_x, \ \tilde{p}_z$, we obtain the corresponding Boyer-Lindquist components $p_r, \ p_{\theta}$ from a direct change of coordinates.

	\item With the initial values of $p_{\mu}$, we calculate the constants of motion:
	\begin{eqnarray}
		\label{eq:energy}
		E&=&-p_t c-q A_t, \\
		L&=&p_{\phi}+q A_{\phi}, \\
		\mathcal{K}&=&p_\theta^2+\cos^2\theta\left[a^2\left(m^2-\frac{E^2}{c^4}\right)+\frac{L^2}{\sin^2\theta}\right],
	\end{eqnarray}
and the combination
	\begin{equation}
		\mathcal{Q}=\mathcal{K}+\left(L-\frac{aE}{c^2}\right)^2.
	\end{equation}
The mass of the particle is the fourth constant of motion, that we calculate as a check for consistency:
	\begin{equation}
		m=\sqrt{\frac{-p^{\mu}p_{\mu}}{c^2}}.
	\end{equation}

	\item We proceed to integrate Eqs. (\ref{eq:drdlambda})-(\ref{eq:dtdlambda}) numerically. To that aim, we apply a fourth order Runge-Kutta subroutine.

	\item Based on the expression of \citet{crawford2002}, we calculate the local Lorentz factor $\gamma$ as measured by a Zero Angular Momentum Observer $u_{\mu} \rightarrow \left(u_t,\vec{0} \right)$.

\end{enumerate}

\section{Results}
\label{s:results}
\begin{figure}
  \centering
    \includegraphics[angle=270,width=.9\columnwidth]{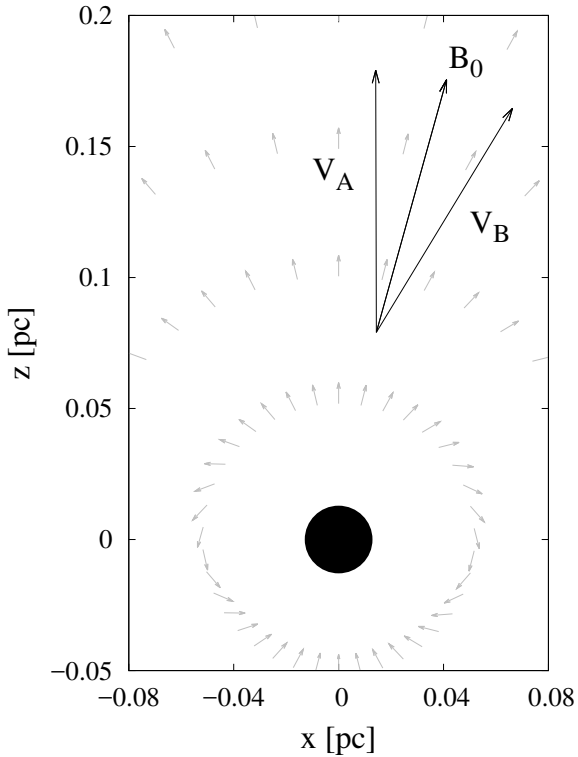}
	\caption{Initial disposition of the velocities of particles for cases A and B with respect to the initial gravito-magnetic field line. Because of gravito-magnetic forces, we expect opposite signs for the angular velocity $\omega_{\phi}$.}
	\label{fig:2}
\end{figure}

	We apply the numerical code described in the previous section to the supermassive black hole in M87. We set $M=6\times 10^{9} M_{\odot}$ and $a=0.9 G_{\mathrm{N}} M/c$, as estimated by \citet{gebhardt2011} and \citet{li2009}, respectively.\par

	From observations reported by \citet{broderick2015} we know that the radius of M87* is, at most, $8G_{\mathrm{N}}M/c^2$. This estimation implies an upper limit for the parameter $M_0$ and, correspondingly, for $\alpha$ (see Eq. \ref{eq:eventhorizon}):
\begin{equation}
\label{eq:m0restr}
	M_0 \lesssim 10^{11} M_{\odot}.
\end{equation}

	In the first run, we fix $M_0=10^{11} M_{\odot}$ and sample the values $\kappa_1=10^2 \sqrt{\alpha G_{\mathrm{N}}}$, $\kappa_2=10^3 \sqrt{\alpha G_{\mathrm{N}}}$, and $\kappa_3=10^4 \sqrt{\alpha G_{\mathrm{N}}}$, where $\sqrt{\alpha G_{\mathrm{N}}}$ is Moffat's original prescription for $\kappa$. We set the mass $m=1 \mathrm{g}$ and the initial position $r_0=140 G_{\mathrm{N}}M/c^2$, $\theta_0=0.18$, $\phi_0=0$. For the initial Lorentz factor we use $\gamma=2$. Such parameters are based on recent observational results \citep{mertens2016}.

	We explore different values for the ejection angle: $\theta^{\mathrm{A}}_{\mathrm{ej}}=0$ and $\theta^{\mathrm{B}}_{\mathrm{ej}}=0.3$, which we refer as \emph{case A} and \emph{case B}, respectively. In Fig. \ref{fig:2} we show the disposition of the initial velocities, with the local gravito-magnetic field. Because of gravito-magnetic forces, we expect opposite signs in the angular velocity $\omega_{\phi}$ for each case.\par

\begin{figure}
  \centering
   \includegraphics[angle=270,width=.9\columnwidth]{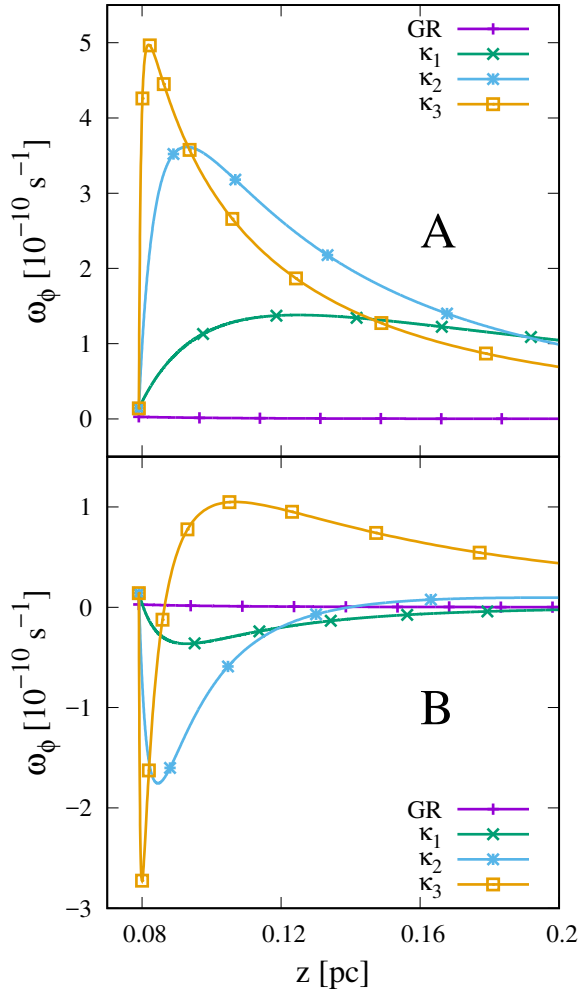}
	\caption{\textit{Top:} Angular velocity $\omega_{\phi}$ as a function of $z$ for case A. Maxima grows with $\kappa$ as a consequence of the increase of gravito-magnetic forces. Rapid decay happens because particles deviate in $\theta$ and get aligned with the field lines, nullifying gravito-magnetic forces. \textit{Bottom:} Angular velocity $\omega_{\phi}$ as a function of $z$ for case B. Initially, $\omega_{\phi}$ is negative due to gravito-magnetic forces. Such forces are absent in GR, where $\omega_{\phi}>0$ due to frame dragging effects. The subsequent behavior of $\omega_{\phi}$ is related to the disposition of gravito-magnetic field lines and the velocity of the particles along the trajectory.}
	\label{fig:3}
\end{figure}

	In Fig. \ref{fig:3} we plot $\omega_{\phi}$, defined as the ratio between $d\phi/d\lambda$ and $dt/d\lambda$, as a function of $z$. We find significant deviations from GR. In case A, rotation along $\phi$ is enhanced by gravito-magnetic forces, leading to higher maxima. On the contrary, for case B, we obtain negative values for $\omega_{\phi}$. This is because gravito-magnetic forces are now directed towards $-\phi$.\par

\begin{figure*}
\centering	
  \includegraphics[angle=270,width=1.5\columnwidth]{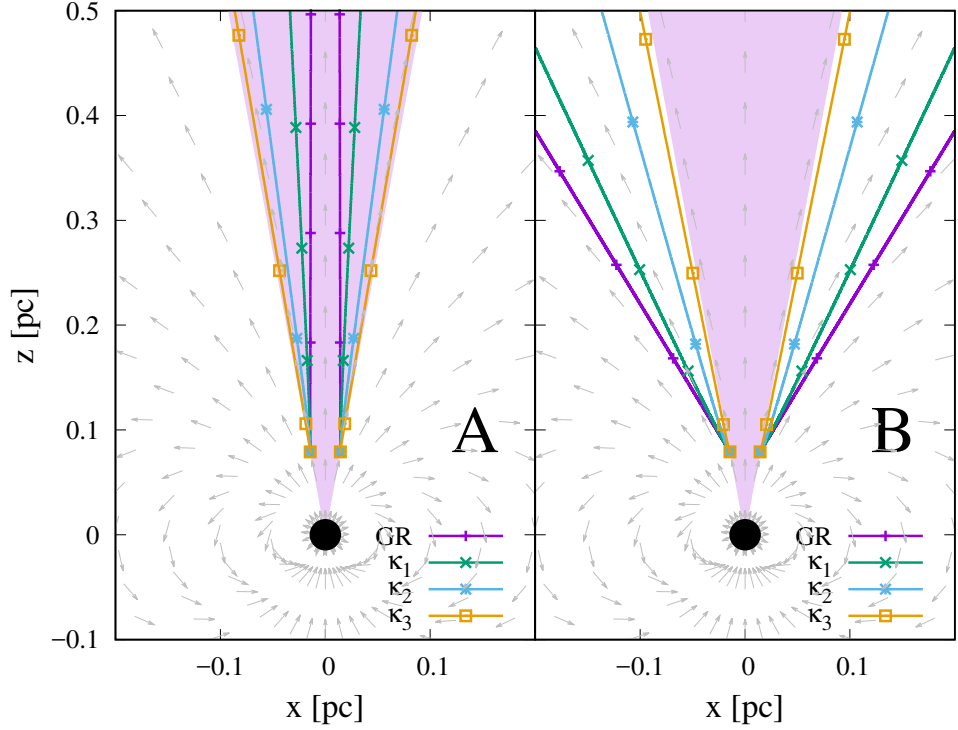}
	\caption{\textit{Left:} $x-z$ projection of trajectories for distinct values of $\kappa$, for particles initially ejected along $z$ (case A). Gravito-magnetic forces leads to deflection in $\theta$. \textit{Right:} $x-z$ projection of trajectories for distinct values of $\kappa$, with initial ejection angle $\theta^{\mathrm{B}}_{\mathrm{ej}}=0.3$ (case B). Gravito-magnetic forces contribute to jet collimation, deviating the particle towards the rotation axis. The filled regions represent the sub-parsec relativistic jet of M87.}
	\label{fig:4}
\end{figure*}

	Along the trajectories, the gravito-magnetic field lines rotate and change their disposition with respect to the velocity of the particle. Then, even for case B, $\omega_{\phi}$ grows for larger $z$. This can be seen in Fig. \ref{fig:4} where we plot the $x-z$ trajectories for both cases, along with gravito-magnetic field lines. The filled region in the latter figure is the relativistic jet of M87*, with opening angle $\Theta \sim 0.18$ \citep{mertens2016}.\par

\begin{figure}
	\centering
    \includegraphics[angle=270,width=\columnwidth]{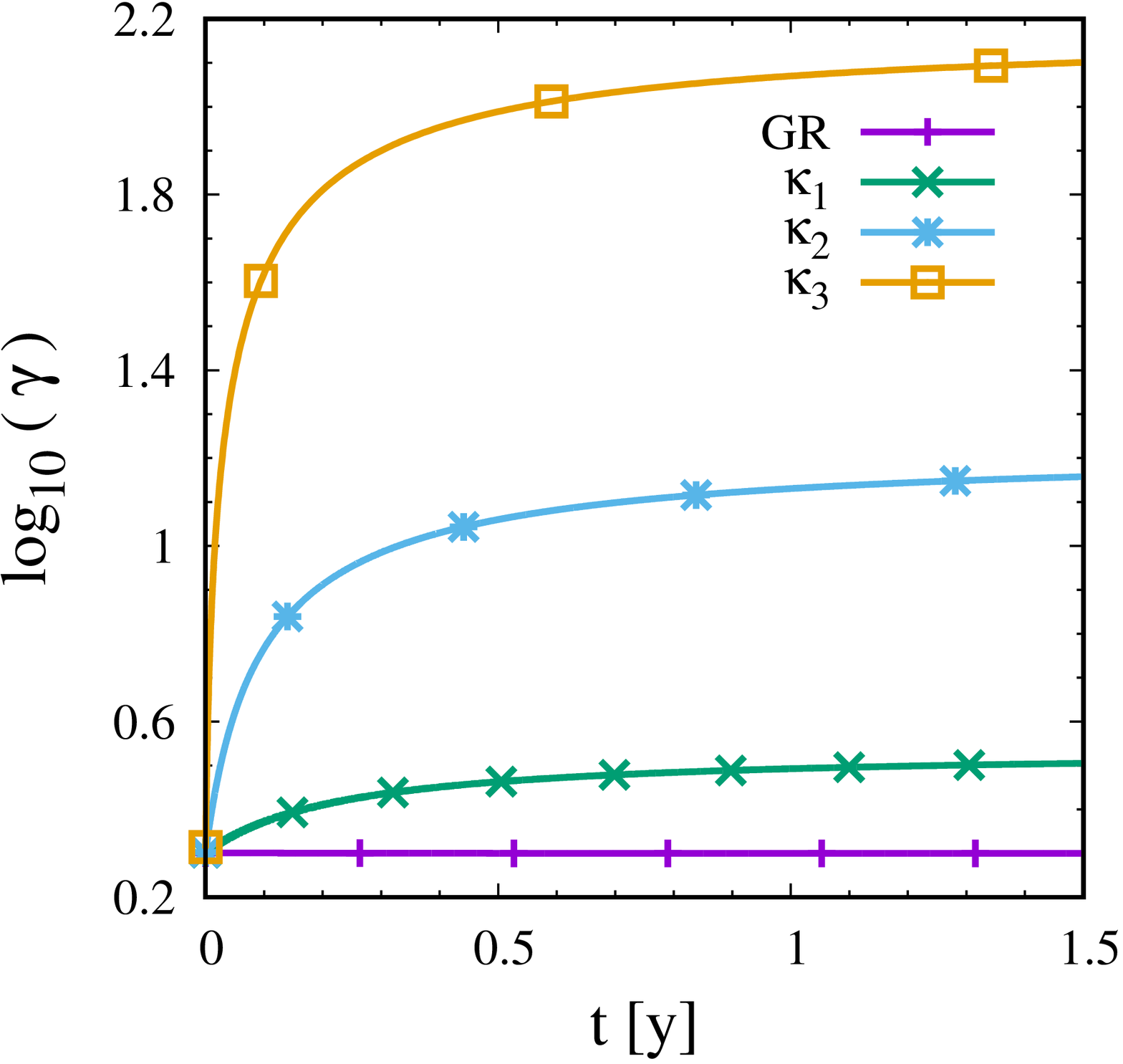}
	\caption{Local Lorentz factor $\gamma$ for particles in case A as a function of $t$ for different values of $\kappa$. The enhancement of repulsive gravito-electric forces with $\kappa$ leads to growing $\gamma$, i.e. particles are gravitationally accelerating. We make use of this fact to state an upper limit for $\kappa$. Particles in case B present similar behaviors for $\gamma$.}
	\label{fig:5}
\end{figure}

	The enhanced gravito-electrical repulsion by growing $\kappa$ affects significantly the kinematic properties of the test particle. In Fig. \ref{fig:5} we see that the local Lorentz factor $\gamma$ grows with time and reaches high values. Then, in the strong field regime of STVG, particles gravitationally accelerate. The energy source for such acceleration is the potential energy term of Eq. (\ref{eq:energy}). This can be used to invoke an active role of gravity in the acceleration of the jet. \par

	However, there are observational constraints on the velocities of the inner jet of M87. The highest value of $\gamma$ estimated by \citet{mertens2016} corresponds to the spine of the jet and is $\gamma\sim10$. Then, based on Fig. \ref{fig:5}, we state the upper limit:
\begin{equation}
	\kappa\leq 10^2 \sqrt{\alpha G_{\mathrm{N}}}.
\end{equation}

	From Fig. \ref{fig:4} we can also notice the effects of gravito-electric and magnetic forces deflecting particles in $\theta$. The repulsive gravito-electric forces accelerate particles in the radial direction, moving them away from the rotational axis in case A. On the contrary, particles move towards the rotational axis in case B. This effect is also facilitated by gravito-magnetic forces since, for instance in case A, when particles acquire positive angular velocity $\omega_{\phi}$, a second order gravito-magnetic force is generated in the polar direction. This results in the motion of particles away from the rotational axis. On the other hand, for particles in case B with negative $\omega_{\phi}$, the second order gravito-magnetic force is directed towards the rotational axis. Through this effect, gravito-magnetic forces could considerably contribute to collimation at the base of the jet, since gravito-magnetic field lines are almost vertical there.

	Now, in a second run, we sample $M_0=10^{10}M_{\odot}$, $10^{11}M_{\odot}$ and $10^{12}M_{\odot}$. The latter violates restriction (\ref{eq:m0restr}) but we include it for consistency checks. Such values for $M_0$ imply the approximate values $\alpha \approx 4$, $13$, and $40$. Notice that, within this values, is included $\alpha\sim 9$ as determined by \cite{moffat2013} and frequently used in references. We take Moffat's weak field limit prescription $\kappa=\sqrt{\alpha G_{\mathrm{N}}}$, and we set $\theta_{\mathrm{ej}}=0$, i. e. particles are ejected along the axis $z$.\par

\begin{figure}
	 \centering
   \includegraphics[angle=270,width=\columnwidth]{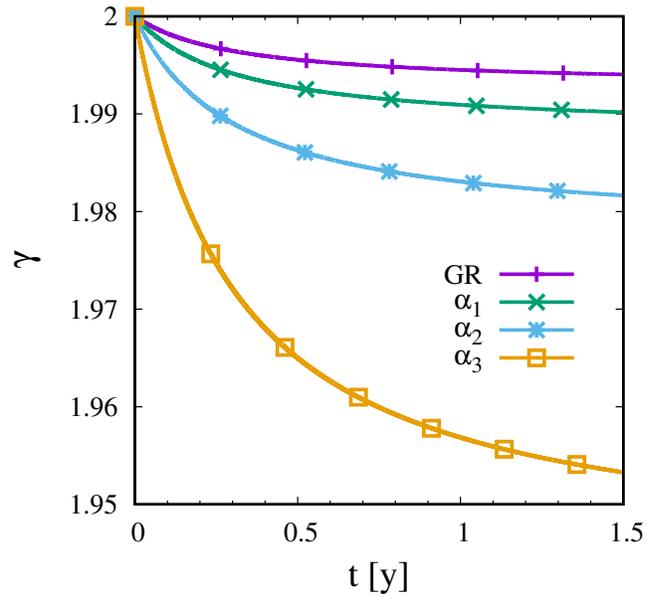}
	\caption{Local Lorentz factor $\gamma$ for different values of $\alpha$, as a function of  $t$. Although Moffat's prescription for $\kappa$ reproduce GR predictions in the Solar System, we can see that this is not the case in the strong field regime. The dependence of curvature with $\alpha$ is highly non-linear, and the values of $\gamma$ decrease deeper for larger $\alpha$. }
\label{fig:6}
\end{figure}

	We find Lorentz-like forces to be negligible and we associate this fact to the small value of $\kappa$. The trajectories, indeed, are almost indistinguishable. However, we find large deviations on kinematic properties. In Fig. \ref{fig:6} we plot the local Lorentz factor $\gamma$ as a function of time, and find that the decrement is greater for larger $\alpha$. Then, although with Moffat's prescription for $\kappa$ repulsion and attraction grow in equal proportion, attraction prevails because the dependence of curvature with $\alpha$ is highly non-linear.

	All these results show that STVG theory has an important impact on the physics of relativistic jets. In the next section, we discuss some interesting applications to jet phenomenology.

\section{Discussion}
\label{s:discussion}
\begin{figure}\centering
  \includegraphics[angle=270,width=\columnwidth]{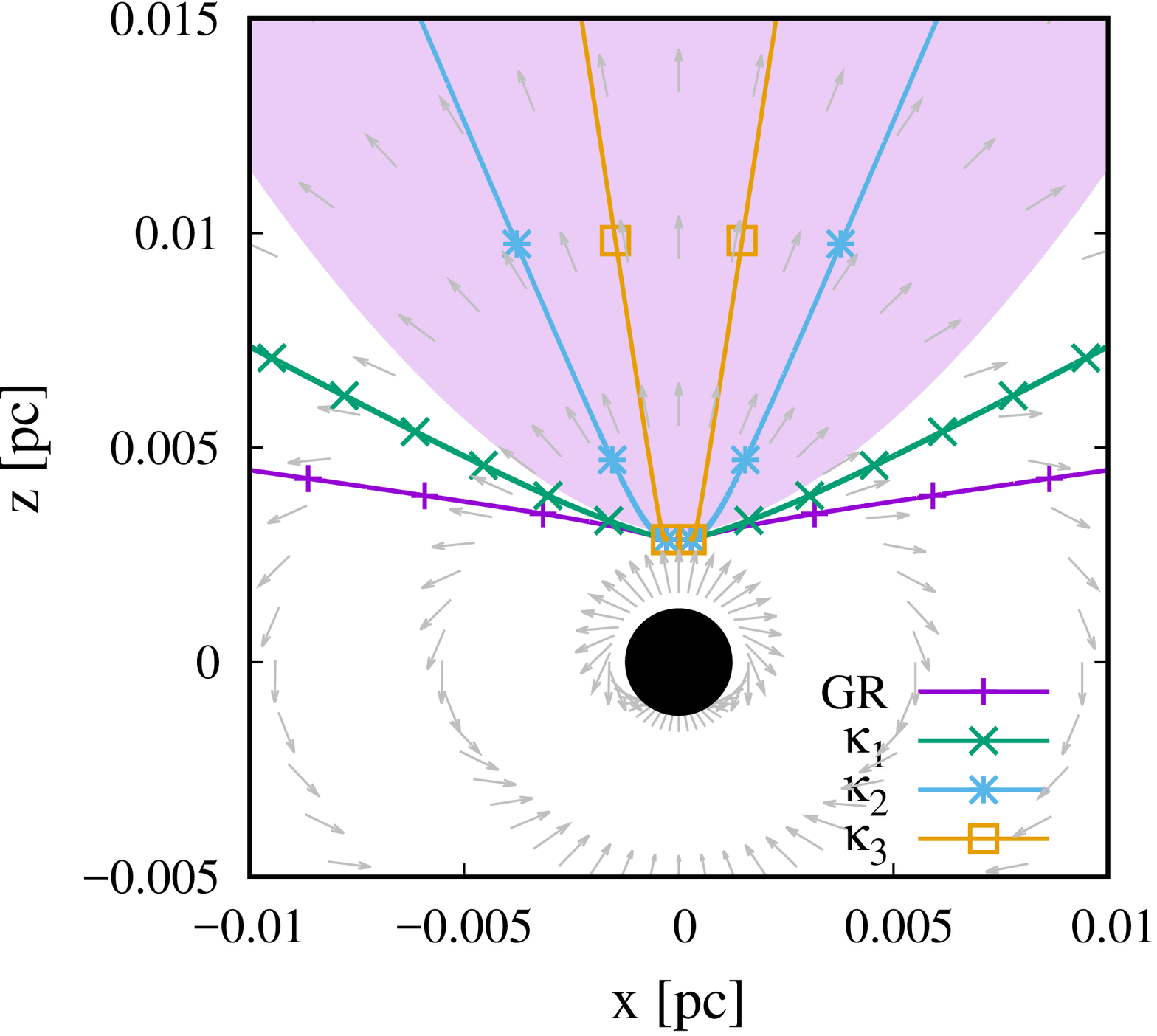}
	\caption{$x-z$ projection of trajectories, for different values of $\kappa$, at the lunching region of the jet. The filled region is the jet in M87, as parametrized by \citet{mertens2016}. Gravito-magnetic forces contribute to the collimation of the jet.}
	\label{fig:7}
\end{figure}

\begin{figure}\centering
  \includegraphics[angle=270,width=\columnwidth]{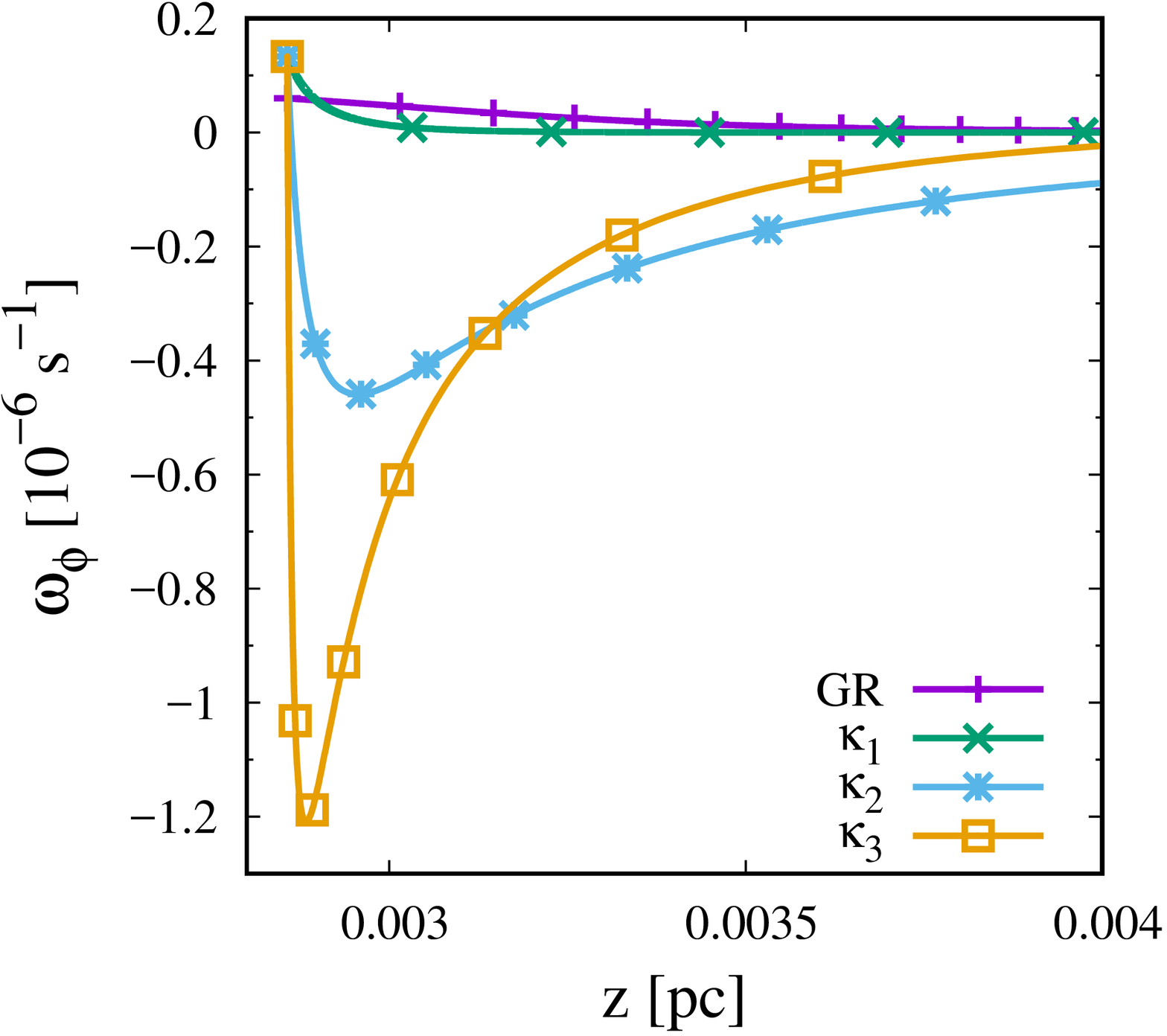}
	\caption{Angular velocity $\omega_{\phi}$ for particles ejected with a wide angle, at the launching region of the jet. The behavior of $\omega_{\phi}$ is related to the disposition of gravito-magnetic field lines and the velocity of the particles along the trajectory. These effects are absent in GR, where particles rotate due to frame-dragging effects.} 
	\label{fig:8}  
\end{figure}

	We have studied the trajectories of particles in STVG-Kerr spacetime. We found that STVG is not equivalent to GR in the strong field regime. In the face of current problems on  the models of jet formation, some STVG predictions seem attractive.\par

	The case of M87 and its jet is particularly interesting since the jet has been recently resolved on scales of 100-1000 Schwarzschild radii \citep{mertens2016}. Very Long Base Line radio observations at 43 GHz have reveled a jet that initially expands with a parabolic profile \citep{asada2012} and then transits to a conical jet at a projected distance of $\sim350$ mas  (2 mas $\approx$ 0.16 pc). The radius of the jet evolves with the distance to the central source as $r_{\rm jet}\propto z^{0.6}$, with significant oscillations that might reflect the growing of Kelvin-Helmholtz instabilities. The observations revealed the existence of a structured jet with clear stratification: a slow outer component and a faster relativistic spine \citep{mertens2016}.

	The jet of M87 is the first one where rotation has been directly observed. The jet first rotates clockwise and then the outer components rotates counterclockwise. Assuming conservation of the specific energy and angular momentum, and assuming Keplerian motion in the accretion disks, a rotation angular velocity of $\omega_{\phi}\sim 10^{-6}$ s$^{-1}$ is obtained \citep{mertens2016}.\par

	At the launching region, the effects of gravito-magnetic forces of STVG are critical. Then, we conjecture that the observed rotation might result from gravitational forces. In order to check the viability of this conjecture, we run our code adopting $r_0=5 R_{\mathrm{S}}$, $M_0=10^{10} M_{\odot}$, $\kappa_1= 10^1 \sqrt{\alpha G_{\mathrm{N}}}$, $\kappa_2=10^2 \sqrt{\alpha G_{\mathrm{N}}}$, $\kappa_3=10^3 \sqrt{\alpha G_{\mathrm{N}}}$, and a wide ejection angle, as expected from the Blandford-Payne mechanism for jet launching \citep{blandford1982,spruit2010}. The $x-z$ trajectories obtained for different values of $\kappa$ are shown in Fig. \ref{fig:7}. The filled region in the latter figure is the jet, as parametrized by \citet{mertens2016} on this scale. We notice the effects of gravito-magnetic and electric forces contributing to the collimation of the jet.

In Fig. \ref{fig:8} we plot the angular velocity $\omega_{\phi}$ as a function of $z$, for different values of $\kappa$. We can see that the initial gravito-magnetic force leads to counter-rotation in $\phi$. But, as we mentioned in the previous section, the field lines rotate along the trajectory and, from a given $z$, the sign of gravito-magnetic forces change and $\omega_{\phi}$ starts growing. The scale where jet rotation gets inverted, and the order of magnitude for $\omega_{\phi}$, are consistent with the observational results and the phenomenological modeling of \citet{mertens2016}.

The standard magnetic model for jet formation has contradictory conditions for strong jet collimation and strong acceleration, since they require distinct inclination angles for the magnetic field lines. It is usually argued that collimation might be produced by some external agent. For instance, \citet{spruit1997} propose a collimation mechanism based on a dipole-like magnetic field. Since the gravito-magnetic field of STVG is independent of the standard magnetic field, it may serve as such external agent as well. 

	Our discussion suggests that gravity, through STVG, may play an important role in the formation of astrophysical jets. We should mention, however, that similar statements have been made for GR. For instance, \citet{defelice1972} studied the allowed ranges of variation for the coordinate $\theta$ in the geodesics of Kerr spacetime. They found a set of geodesics for unbound particles, which they called \emph{vortical orbits}, that spiral around the symmetry axis and never cross the equatorial plane. Further, \citet{defelice1992} showed that perturbing particular vortical orbits leads to collimation around the symmetry axis.\par

	In order to find out whether STVG is more adequate than GR to model jet formation, we analyze the amount of vortical orbits in STVG-Kerr spacetime. We adapt the conditions \citet{defelice1972} for vortical orbits to the modified equation of motion (\ref{eq:dthetadlambda}) and find:
\begin{eqnarray}
\label{eq:vortical1}
\Gamma&>&0, \\
\label{eq:vortical2}
-a^2\Gamma &\leq& \mathcal{Q}+L^2\leq a^2\Gamma,\\
\label{eq:vortical3}
L^2+\mathcal{Q}&\leq& L^2 \leq \frac{\left( a^2 \Gamma+L^2+\mathcal{Q}\right)^2}{4a^2\Gamma},
\end{eqnarray}
where $\Gamma=E^2/c^4-m^2$. We vary the initial angle $\theta_0$, and the ejection angle $\theta_{\mathrm{ej}}$, and test whether the resulting trajectories satisfy the latter vortical conditions.\par

\begin{figure}[h]\centering
  \includegraphics[angle=270,width=\columnwidth]{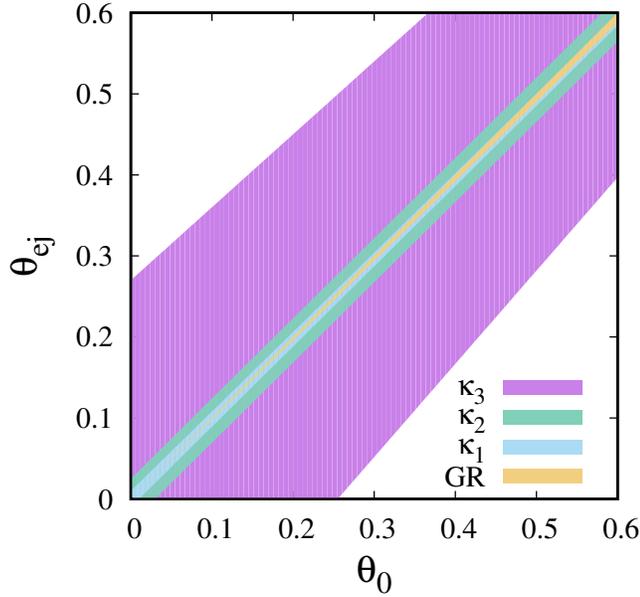}
	\caption{Parameter space $\theta_0-\theta_{\mathrm{ej}}$, where regions of vortical orbits has been filled. STVG includes a larger amount of vortical orbits for growing $\kappa$, as compared to GR. Therefore, STVG seems more suitable for explaining jet formation.}
	\label{fig:9}  
\end{figure}

 In Fig. \ref{fig:9} we plot the parameter space $\theta_0-\theta_{\mathrm{ej}}$, and fill the regions that include vortical orbits. As we can see, the number of vortical orbits grows with $\kappa$. This occurs because gravito-magnetic forces led to better collimation and gravito-electrical repulsion enhances radial acceleration. Future work will be devoted to the analysis and perturbation of such orbits.

\section{Conclusions}
\label{s:conclusions}
	We applied STVG theory to the black hole and jet in M87. We followed Moffat's prescription and approximated the scalar field $G$ as a constant. Also,  we approximated $m_{\phi}=0$ because its effects manifest at kiloparsec scales, and we were interested on sub-parsec structures. We found resemblances of this regime with Einstein-Maxwell formalism.\par

	We described STVG-Kerr spacetime. The black hole event horizon and ergosphere grow in size with the free parameter $\alpha$. Since there are constraints on the size of M87* from mm-VLBI observations, we set an upper limit for the related parameter $M_0$ of the theory.\par

	Unlike many gravitational theories, STVG is not purely geometrical. Instead, it includes a Yukawa-type vector field $\boldsymbol{\phi}$ that couples to matter. We characterized the effects of such vector field on the motion of particles in STVG-Kerr spacetime. Repulsive gravito-electrical components counteracts enhanced attraction and serves to recover classical limits, while gravito-magnetic components involve novel predictions of STVG.\par

	We derived the equations of motion for test particles in STVG-Kerr spacetime. Such equations depend on the coupling constant $\kappa$. Moffat proposed the value $\kappa=\sqrt{\alpha G_{\mathrm{N}}}$ for recovering classical limits, but this prescription only works on the weak field regime. Instead, we treated $\kappa$ as a free parameter, and study its effects on particle motion.\par

	We developed a code that integrates the trajectories of particles in a relativistic jet, and used it to model the jet in M87. First, we used the code to sample the parameter $\kappa$. The effects of gravito-magnetic forces arose and the theory clearly deviates from GR. Because of gravito-electrical repulsion, we found that particles gravitationally accelerate and reach high Lorentz factors. Based on observational constraints for velocities in the relativistic jet of M87, we determinated an upper limit for $\kappa$ in our model. On the other hand, gravito-magnetic forces influenced the angular velocity $\omega_{\phi}$, depending critically on the ejection angle. As a third effect, we found collimation and de-collimation in the coordinate $\theta$, also depending on the initial ejection angle.\par

 Then, we sampled $\alpha$, adopting the prescription of Moffat for the parameter $\kappa$. The effects of Lorentz-like forces on trajectories resulted negligible. However, the increase of the energy of the black hole with $\alpha$ led to a larger decrease of the particle velocity, as compared with GR.\par

 From both runs, we concluded that STVG differs with GR not only far from the gravitational source, where phenomena associated with dark matter use to happen, but also in the strong field regime.\par

	We compared observational results on the formation zone of the jet in M87 with predictions of STVG. We concluded that gravity, through STVG, might play an important role in the process of acceleration and collimation of the jet. This conclusion is supported by the analysis of vortical orbits in STVG. Interestingly enough, we found that the observed rotation and counter-rotation of the jet in M87 could be a consequence of the gravito-magnetic field.

\acknowledgments This work was supported by grants AYA2016-76012-C3-1-P (Ministro de Educaci\'on, Cultura y Deporte, Espa\~na) and PIP 0338 (CONICET, Argentina). We would like to thank Luciano Combi and Santiago del Palacio for helpful discussions.

\bibliographystyle{spr-mp-nameyear-cnd}
\bibliography{biblio}

\end{document}